\documentclass[8.5pt,twoside,twocolumn]{article}
\usepackage[super,sort&compress,comma]{natbib}
\usepackage{times,mathptmx}
\usepackage{balance} 

\usepackage{graphicx} 
\usepackage{lastpage}
\usepackage[format=plain,justification=raggedright,singlelinecheck=false,font=small,labelfont=bf,labelsep=space]{caption} 

\usepackage{amsmath}
\usepackage{amssymb}
\usepackage{amsthm}
\usepackage{mathtools}
\usepackage[np,autolanguage]{numprint}
\usepackage[normalem]{ulem}
\usepackage{color}
\usepackage{setspace}
\usepackage{enumerate}
\usepackage{booktabs}
\usepackage{bm}
\DeclareGraphicsExtensions{.pdf, .jpg, .tif,.png}
\usepackage{acronym}
\acrodef{DNA}[DNA]{Deoxyribonucleic acid}
\acrodef{SAW}[SAW]{standing acoustic wave}
\acrodef{PDMS}[PDMS]{polydimethylsiloxane}
\acrodef{DOF}[DOF]{degrees of freedom}

\DeclareMathOperator{\rp}{Re}

\newcommand{\bv}[1]{\mathbf{#1}}

\newcommand{\I}{\ensuremath{\imath}}
\newcommand{\scdot}{\ensuremath{\!\cdot\!}}
\newcommand{\figpath}[1]{#1}

\newcommand{\umu}{\ensuremath{\,\mu\text{m}}}

\begin{document}
\renewcommand{\thefootnote}{}
\twocolumn[
  \begin{@twocolumnfalse}
\noindent\LARGE{\textbf{%
Investigation of acoustic streaming patterns around oscillating sharp edges\dag%
}
}
\vspace{0.6cm}

\noindent\large{\textbf{%
Nitesh Nama,\textit{$^{a}$}
%
Po-Hsun Huang,\textit{$^{a}$}
%
Tony Jun Huang,$^{\ast}$\textit{$^{ab}$}
%
and
Francesco Costanzo$^{\ast}$\textit{$^{ac}$}%
}}\vspace{0.5cm}

%

\noindent \normalsize{%
\textbf{Abstract}\qquad
Oscillating sharp edges have been employed to achieve rapid and homogeneous mixing in microchannels using acoustic streaming. Here we use a perturbation approach to study the flow around oscillating sharp edges in a microchannel. This work extends prior experimental studies to numerically characterize the effect of various parameters on the acoustically induced flow. Our numerical results match well with the experimental results. We investigated multiple device parameters such as the tip angle, oscillation amplitude, and channel dimensions. Our results indicate that, due to the inherent nonlinearity of acoustic streaming, the channel dimensions could significantly impact the flow patterns and device performance.%
}
\vspace{0.5cm}
 \end{@twocolumnfalse}
]

\footnotetext{\dag Submitted for publication in \emph{Lab on a Chip}, 12 September 2014.}
\footnotetext{$^{a}$Department of Engineering Science and Mechanics,
The Pennsylvania state University,
University Park,
PA 16802, USA.
E-mail: junhuang@psu.edu}
\footnotetext{$^{b}$Department of Bioengineering,
The Pennsylvania state University,
University Park,
PA 16802, USA.}
\footnotetext{$^{c}$Center for Neural Engineering,
The Pennsylvania state University,
University Park,
PA 16802, USA.
E-mail: costanzo@engr.psu.edu}
\footnotetext{$^{\ast}$Corresponding authors: F.~Costanzo, E-mail: costanzo@engr.psu.edu; T.~J.~Huang, E-mail: junhuang@psu.edu.}

\section{Introduction}
\label{section: introduction}
Microfluidic devices can be effective in many applications including biomedical diagnostics, drug delivery, chemical synthesis, and enzyme reactions\cite{Yu2014Microfluidic-Bl-0,Lee2013The-Third-Decad-0}. An important requirement for these systems is the ability to rapidly and efficiently mix small amounts of samples at microscales\cite{Coleman2005A-Sequential-In-0,Wang2011Mixing-Enhancem-0}. Various techniques have been utilized to enable rapid mixing in microfluidic devices including chaotic advection \cite{Stroock2002a,Chen2004,LEE2007,Lin2007,Che2011Analysis-of-Cha-0,Hsu2006Spatio-Temporal-0}, hydrodynamic focusing \cite{Park2006, Mao2007a}, electrokinetically driven mixing \cite{Lee2004,Harnett2008,Sigurdson2005,Ng2009,Coleman2006}, 3D combinatorial bubble-based mixing\cite{Neils2004,Lim2011}, and thermally- \cite{Tsai2002,Xu2010Thermal-Mixing--0} as well as optically-induced \cite{Hellman2007} mixing.  Recently, acoustic-based mixers \cite{Lin2012,Shi2008,Rezk2012,Tseng2006,Luong2011High-Throughput-0} have generated significant interest because of their non-invasive nature. These mixers utilize acoustic waves to perturb the laminar flow pattern in microchannels to achieve rapid and homogeneous mixing\cite{Wixforth2004,Frommelt2008,Sritharan2006,Yu2006,Yang2001,Oberti2009}. In particular, acoustically driven, oscillating bubbles have been used to achieve fast and homogeneous mixing by generating acoustically-induced microvortices \cite{Ahmed2009,Collin2006Laminar-Flow-Mi-0,Liu2002Bubble-Induced--0}. Bubble-based acoustic mixers have been utilized for enzyme reaction characterization \cite{Xie2012}, \acs{DNA} hybridization enhancement \cite{Liu2002,Rodaree2011}, chemical gradient generation \cite{Ahmed2013}, and optofluidic modulators \cite{Huang2012, Huang2013,Hashmi2012Oscillating-Bub-0}. However, bubble-based acoustic mixers have also proven to be somewhat challenging due to bubble instability, heat generation, and hard-to-control bubble-trapping processes\cite{Huang2013}. To overcome these difficulties, we recently reported a \emph{sharp-edge}-based micro-mixer \cite{Huang2013} where the flow field is perturbed using microvortices generated by an acoustically oscillating sharp edge. The performance of the sharp-edge based micro-mixer was found to be very close to that of the bubble-based micro-mixer with the added advantage of convenient and stable operation over bubble-based micro-mixers. However, to realize the full potential of these devices and explore further applications, a deeper understanding of the flow field around oscillating sharp edges is required.

Steady streaming around obstacles in an oscillating incompressible fluid has been studied extensively \cite{Riley1998, Riley2001, Lighthill1978, Mathematics1973,Bruus2011Forthcoming-Lab-0}. Due to the dissipative nature of the fluid, the response to a time-harmonic forcing is generally not harmonic. The fluid's response to a harmonic forcing can be viewed as a combination of a time harmonic response, generally referred to as acoustic response, and a remainder, referred to as acoustic streaming \cite{Ding2013}. Time averaging of the Navier-Stokes equations yields a term analogous to the Reynolds stress (normally observed in turbulent flows) which causes a ``slow'' steady streaming around obstructions in the flow field.  This can also be interpreted by saying that the nonlinear hydrodynamic coupling results in a partial transmission of the acoustic wave energy to the fluid as steady momentum resulting in acoustic streaming \cite{Lieu2012}. Since the latter is a byproduct of the acoustic attenuation due to viscous dissipation, it provides a unique way to utilize the dominant viscous nature of microfluidic flows \cite{Frampton2003}.  While bubble-based mixers have been extensively studied both analytically and numerically \cite{Ahmed2009a, Ahmed2009, Mao2009, Liu2002}, the knowledge of flow fields around oscillating sharp edges is limited. Lieu \emph{et al.} \cite{Lieu2012} have studied the flow around obstacles in an oscillating incompressible flow field.  However, Lieu \emph{et al.}\cite{Lieu2012} do not discuss possible singularities induced in the flow by the geometry of the obstacles.  In addition, their analysis is not directly applicable to our system, which is characterized by acoustic wave propagation, the latter requiring an explicit modeling of the fluid compressibility.  Although steady streaming has been widely studied for cases where the fluid can be treated as infinite, little attention has been given to this phenomenon in confined flows\cite{Lieu2012}.  The basic hydrodynamic traits of low Reynolds number flows in microfluidic channels are dictated by the motion of the walls.  This, coupled with the inherent nonlinearity of the acoustic streaming phenomena, implies that the geometrical dimensions of the microfluidic channel and the boundary conditions significantly impact the flow field around the oscillating sharp edge inside a microchannel.  These considerations justify a numerical approach to study the flows in question in which the geometry of the walls can be accurately represented.

In this work, we numerically investigate the acoustic streaming generated in a fluid by oscillating sharp edges inside a microchannel. We build on our previous experimental studies and aim at characterizing the effect of various parameters on the micro-eddies around the sharp edges. We model the fluid as compressible and linear viscous so that the fluid's equations of motion are the compressible Navier-Stokes equations.  These are intrinsically nonlinear and characterized by different behaviors over wide ranges of time and length scales. The flow on the large length- and time-scales arises from the acoustic excitation at much smaller time and length scales \cite{Frommelt2008}. Consequently, a direct solution of the compressible Navier-Stokes equation remains computationally challenging even with modern computational tools.  To overcome this difficulty, we employ Nyborg's perturbation approach\cite{Nyborg1998Acoustic-Stream0} complemented by periodic boundary conditions. To capture the singularity in the flow field, we refine the mesh near the tip of the sharp edge using an adaptive mesh refinement strategy. Our approach is general in the sense that we do not make a priori assumptions about specific flow regimes in selected regions of the computational domain.  After identifying boundary conditions that lead to predictions matching experimental observations, we numerically investigate the effects of various parameters like tip angle, displacement amplitude, and channel dimensions on the streaming velocity and particle mean trajectories. The numerical methods and results presented in this article will be useful in optimizing the performance of acoustofluidic devices and providing design guidelines.

\section{Micro-acousto-fluidic channel with sharp edges}
\label{sec:Experimetal Procedure}
\begin{figure}[htb]
    \centering
    \includegraphics[width=8.5cm]{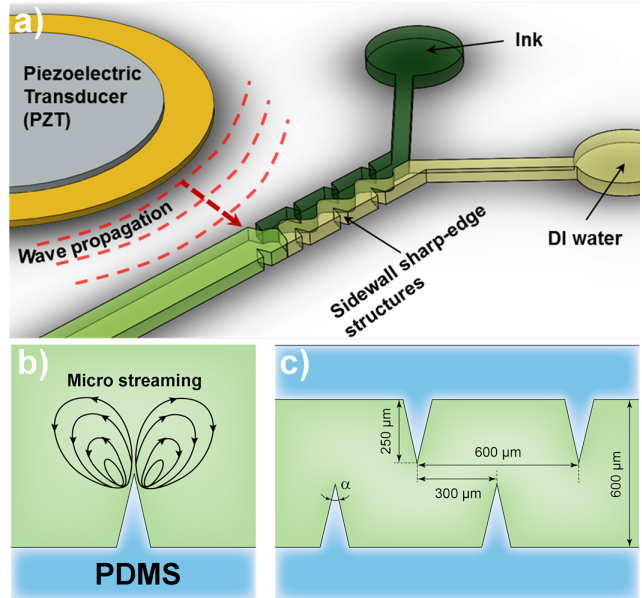}
    \caption{(a) Schematic of the device showing a micro-fluidics channel with sharp-edge structures on its side walls.  The channel walls are subjected to a time-harmonic excitation produced by a piezoelectric transducer placed on one side of the channel.  (b) Typical micro streaming patterns produced in the fluid occupying the channel as a response to piezoelectric excitation.  (c) Typical geometric dimensions of the corrugated channel.}
    \label{fig: Schematic}
\end{figure}

\begin{figure}[htb]
    \centering
    \includegraphics[width=8.5cm]{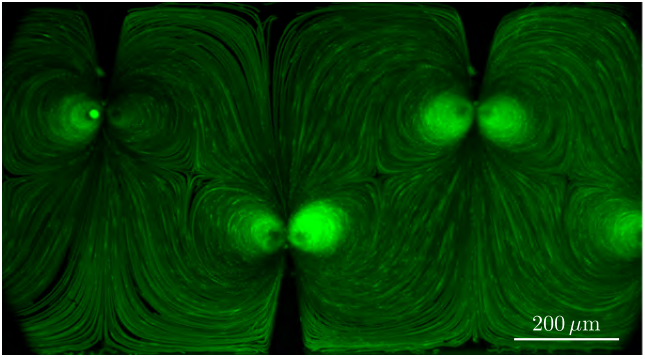}
    \caption{Experimentally observed trajectories of $1.9\umu$ diameter fluorescent beads in our acoustically oscillated micro-mixer with sharp edges.  The geometry of the micro-channel is described in Fig.~\ref{fig: Schematic}(c) except for the fact that here, the tips of the sharp edges are $200\umu$ from the wall instead of $250\umu$.  The driven oscillation is harmonic with a frequency equal to $\np[kHz]{4.75}$.}
    \label{fig: ExperimentalTraces}
\end{figure}

Figure~\ref{fig: Schematic}(a)
provides a schematic of the oscillating sharp-edge-based acoustofluidic micromixer.  A single-layer \ac{PDMS} channel with eight sharp-edges on its sidewall (four on each side) was fabricated using standard soft lithography and bonded onto a glass slide. A piezoelectric transducer (model no.~273-073, RadioShack\textsuperscript{\textregistered}) was then attached adjacent to the \ac{PDMS} channel. Upon the actuation of the piezoelectric transducer, the sharp-edges were acoustically oscillated with a frequency of $\np[kHz]{4.75}$.  These oscillations generate a pair of counter-rotating vortices (double-ring recirculating flows) in the fluid around the tip of each sharp-edge, as shown in Fig.~\ref{fig: Schematic}(b). Typical channel dimensions are indicated in Fig.~\ref{fig: Schematic}(c).  To visualize and characterize the streaming flow inside the channel, a solution containing $1.9\umu$ diameter dragon green fluorescent beads (Bangs Laboratories, Inc.\texttrademark) was introduced into the channel.  The typical bead trajectories observed in experiments are shown in Fig.~\ref{fig: ExperimentalTraces}.

\section{Modeling}
\label{subsec: Numerical Modeling}
We denote vector quantities by boldface type and scalars by a normal-weight font.

\begin{figure}[htb]
    \centering
    \includegraphics[width=8.5cm]{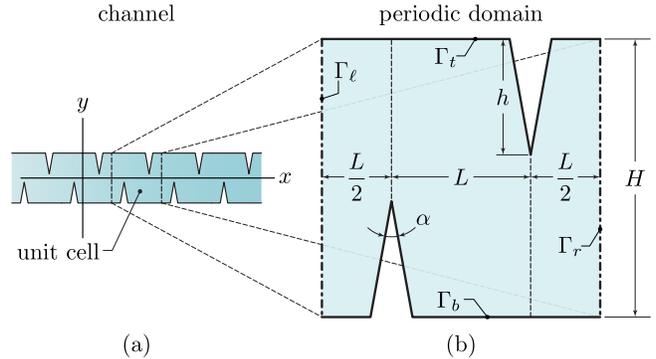}
    \caption{(a) A portion of the microfluidics device. (b) Definition of a periodic cell forming the device.
}
\label{fig: periodic cell def}
\end{figure}

The mass and momentum balance laws governing the motion of a linear viscous compressible fluid are \cite{GurtinFried_2010_The-Mechanics_0,Sadhal2012, Ding2013}
\begin{gather}
\label{Eq: continuity}
\frac{\partial \rho }{\partial t}+ \nabla \scdot (\rho \bv{v})=0 
\shortintertext{and}
\label{Eq: navierstokes}
\rho \frac{\partial \bv{v}}{\partial t} + \rho (\bv{v} \scdot\nabla) \bv{v} = -\nabla p + \mu \nabla^{2} \bv{v} + (\mu _{b} + \tfrac{1}{3} \mu) \nabla (\nabla \scdot \bv{v}),
\end{gather}
where $\rho$ is the mass density, $\bv{v}$ is the fluid velocity, $p$ is the fluid pressure, $\mu$ and $\mu_{b}$ are the shear and bulk dynamic viscosities, respectively.  The fields $\rho$, $p$, and $\bv{v}$ are understood to be in Eulerian form \cite{GurtinFried_2010_The-Mechanics_0}, i.e., functions of time $t$ and of the spatial position $\bv{x}$ within a chosen control volume.  Equations~\eqref{Eq: continuity} and~\eqref{Eq: navierstokes}, with appropriate boundary conditions and a constitutive relation linking the pressure to the fluid density, allow one to predict the motion of the fluid.  We assume the relation between $p$ and $\rho$ to be linear:
\begin{equation}
\label{Eq: constitutive}
p = c_{0}^{2} \, \rho, 
\end{equation}
where $c_{0}$ is the speed of sound in the fluid at rest. Direct simulation of this non-linear system of equations poses significant numerical challenges owing to the widely separated length (characteristic wave lengths vs.\ the characteristic geometrical dimensions of the microfluidic channels) and time scales (characteristic oscillation periods vs.\ characteristic times dictated by the streaming speed)\cite{Frommelt2008a}. Because of viscous dissipation, the response of the fluid to a harmonic forcing is, in general, not harmonic. The fluid response is generally thought to be comprised of two components: (\emph{i}) a periodic component with period equal to the forcing period, and (\emph{ii}) a remainder that can be viewed as being steady. It is this second component which is generally referred to as the \emph{streaming motion} \cite{Ding2013}. We employ Nyborg's perturbation technique \cite{Nyborg1998Acoustic-Stream0} in which fluid velocity, density, and pressure are assumed to have the following form:
\begin{subequations}
\label{Eq: expansion}
\begin{align}
\bv{v} &= \bv{v}_{0} + \varepsilon \tilde{\bv{v}}_{1} + \varepsilon^{2} \tilde{\bv{v}}_{2} + O(\varepsilon^{3}) + \cdots,\\
p &= p_{0} + \varepsilon \tilde{p}_{1} + \varepsilon^{2} \tilde{p}_{2} + O(\varepsilon^{3}) + \cdots,\\
\rho &= \rho_{0} + \varepsilon \tilde{\rho}_{1} + \varepsilon^{2} \tilde{\rho}_{2} + O(\varepsilon^{3}) + \cdots,
\end{align}
\end{subequations}
where $\varepsilon$ is a non-dimensional smallness parameter. Following K\"{o}ster \cite{Koster2006Numerical-Simul0}, we define $\varepsilon$ as the ratio between the amplitude of the displacement of the boundary in contact with the piezo-electrically driven substrate (i.e., the amplitude of the boundary excitation) and a characteristic length.  We take the $0$-th order velocity field $\bv{v}_{0}$ to be equal to zero thus assuming the absence of an underlying net flow along the micro-channel. Letting
\begin{equation}
\label{Eq: first and second order defs}
\begin{aligned}
\bv{v}_{1} &= \epsilon \tilde{\bv{v}}_{1}, &
p_{1} &= \epsilon \tilde{p}_{1}, &
\rho_{1} &= \epsilon \tilde{\rho}_{1},
\\
\bv{v}_{2} &= \epsilon^{2} \tilde{\bv{v}}_{2}, &
p_{2} &= \epsilon^{2} \tilde{p}_{2}, &
\rho_{2} &= \epsilon^{2} \tilde{\rho}_{2},
\end{aligned}
\end{equation}
substituting Eqs.~\eqref{Eq: expansion} into Eqs.~\eqref{Eq: continuity} and~\eqref{Eq: navierstokes}, and setting the sum of all the terms of order one in $\varepsilon$ to zero, the following problem, referred to as the first-order problem, is obtained:
\begin{gather}
\label{Eq: firstordercontinuity}
\frac{\partial \rho_{1} }{\partial t} + \rho_{0} (\nabla \scdot \bv{v}_{1}) = 0, 
\\
\label{Eq: firstorderNS}
\rho_{0} \frac{\partial \bv{v}_{1}}{\partial t} = -\nabla p_{1} + \mu \nabla^{2} \bv{v}_{1} + (\mu _{b}+\tfrac{1}{3}\mu)\nabla (\nabla \scdot \bv{v}_{1}).
\end{gather}
Repeating the above procedure for the terms of order two in $\varepsilon$, and averaging the resulting equations over a period of oscillation, the following set of equations, referred to as the second-order problem, is obtained:
\begin{gather}
\label{Eq: secondordercontinuity}
\left\langle \frac{\partial \rho_{2}}{\partial t} \right\rangle + \rho_{0} \nabla \scdot \left\langle \bv{v}_{2} \right\rangle = -\nabla \scdot \left\langle \rho_{1} \bv{v}_{1}\right\rangle\!,
\\
\label{Eq: secondorderNS}
\begin{multlined}[b]
\rho_{0} \left\langle \frac{\partial \bv{v}_{2}}{\partial t} \right\rangle + \left\langle \rho_{1} \frac{\partial \bv{v}_{1}}{\partial t} \right\rangle + \rho_{0} \left\langle \bv{v}_{1} \scdot \nabla \bv{v}_{1} \right\rangle
\\
= -\nabla \left\langle p_{2} \right\rangle 
+ \mu \nabla^{2} \left\langle \bv{v}_{2} \right\rangle + (\mu_{b}+\tfrac{1}{3}\mu) \nabla \nabla \scdot \left\langle \bv{v}_{2} \right\rangle\!,
\end{multlined}
\end{gather}
where $\left \langle x \right \rangle$ denotes the time average of the quantity $x$ over a full oscillation time period.  The above sets of equations need to be complemented by appropriate boundary conditions.  Referring to Figs.~\ref{fig: Schematic}(a) and~\ref{fig: periodic cell def}, we observe that the device consists of an assemblage of identical cells.
While the number of cells in a device is finite, we assumed that the flow in each cell is identical to the flow in any other cell, and therefore we used the cell in Fig.~\ref{fig: periodic cell def}(b) as our computational domain with the stipulation that computed flow must satisfy \emph{periodic boundary conditions}.  As discussed later, the solution of the streaming problem under periodic boundary conditions predicted flows that agreed well with those experimentally observed.  The latter also have an anti-symmetric pattern \emph{internal} to the selected unit cell corresponding to the region of width $L$ in Fig.~\ref{fig: periodic cell def}(b).  In these figures we have also indicated the relevant geometric descriptors that define our solution domain as well as the labels identifying specific regions of the boundary.

\subsection{Boundary conditions}
\label{subsec: loading conditions}
In this work, we do not solve a coupled piezo-electro-elastic problem to relate the voltage control on the transducer to corresponding loading conditions on the channel.  Rather, we formulate simple boundary conditions based on observations from our prior experimental work.  Referring to Figs.~\ref{fig: Schematic}(a) and~\ref{fig: periodic cell def}, we observe that the diameter of the transducer is much larger than the transverse width of the channel ($\approx 600\umu$) so that the channel can be assumed to be subject to a plane wave parallel to the $x$ direction and traveling in the $y$ direction.  Hence, we assume that the boundary portions $\Gamma_{t}$ and $\Gamma_{b}$ (the solid lines in Fig.~\ref{fig: periodic cell def}(b)) are subject to a displacement field $\bv{w}(\bv{x},t)$ of the following form
\begin{equation}
\label{eq: displacement boundary condition}
\begin{aligned}
\bv{w}(\bv{x}_{t},t) &= \bv{w}_{c}^{t} \cos(2 \pi f t) + \bv{w}_{s}^{t} \sin(2 \pi f t),
\\
\bv{w}(\bv{x}_{b},t) &= \bv{w}_{c}^{b} \cos(2 \pi f t) + \bv{w}_{s}^{b} \sin(2 \pi f t),
\end{aligned}
\end{equation}
where $\bv{w}_{c}^{t}$, $\bv{w}_{c}^{b}$, $\bv{w}_{s}^{t}$, and $\bv{w}_{s}^{s}$ are vector-valued constants, and where $f$ is the transducer oscillation frequency in hertz.  Consistently with the asymptotic expansion in Eqs.~\eqref{Eq: expansion}, the boundary conditions on $\Gamma_{t}$ and $\Gamma_{b}$ for the first-order problem are\cite{Bradley1996Acoustic-Stream0,Koster2006Numerical-Simul0} $\bv{v}_{1} (\bv{x}_{t,b},t) =  \partial \bv{w}(\bv{x}_{t,b},t)/\partial t$, which gives
\begin{equation}
\label{eq: boundary conditions fop gamma bt}
\bv{v}_{1} (\bv{x}_{t,b},t) = -2\pi f [\bv{w}_{c}^{t,b} \sin(2 \pi f t) - \bv{w}_{s}^{t,b} \cos(2 \pi f t)],
\end{equation}
where the subscripts and superscripts $t$ and $b$ stand for `on $\Gamma_{t}$' and `on $\Gamma_{b}$', respectively.  For the second-order problem we have\cite{Bradley1996Acoustic-Stream0,Koster2006Numerical-Simul0}
\begin{equation}
\label{eq: boundary conditions sop gamma bt}
\bv{v}_{2}(\bv{x}_{t,b},t) = - \left\langle (\bv{w}(\bv{x}_{t,b},t) \cdot \nabla) \bv{v}_{1}(\bv{x}_{t,b},t) \right\rangle.
\end{equation}
As already mentioned, and referring to Figs.~\ref{fig: ExperimentalTraces} and~\ref{fig: periodic cell def}, experimental results are characterized by distinctive symmetries.  Hence, for both the first- and second-order problems, we enforce periodic boundary conditions along the $x$ direction, that is on $\Gamma_{\ell}$ and $\Gamma_{r}$.  Specifically, for all pairs of homologous points $\bv{x}_{\ell}$ and $\bv{x}_{r}$ on $\Gamma_{\ell}$ and $\Gamma_{r}$, respectively, we demand that
\begin{equation}
\label{eq: periodic boundary conditions}
\bv{v}_{1} (\bv{x}_{\ell},t) = \bv{v}_{1} (\bv{x}_{r},t)
\quad \text{and} \quad
\bv{v}_{2} (\bv{x}_{\ell},t) = \bv{v}_{2}(\bv{x}_{r},t).
\end{equation}
As far as the $y$ direction is concerned, we observe that the wave length of the forced oscillations in the substrate is much larger than the channel's width.  Hence, we subject $\Gamma_{t}$ and $\Gamma_{b}$ to identical (uniform) boundary conditions as though the channel were rigidly and harmonically displaced in the vertical direction:
\begin{equation}
\label{eq: periodic boundary conditions bt}
\bv{w}(\bv{x}_{b},t) = \bv{w}(\bv{x}_{t},t).
\end{equation}
For comparison purposes, and noticing that the experiments suggest the presence of an anti-symmetric pattern relative to the vertical mid-line of the solution domain, we have also considered the following boundary condition:
\begin{equation}
\label{eq: anti periodic boundary conditions tb}
\bv{w}(\bv{x}_{b},t) = -\bv{w}(\bv{x}_{t},t).
\end{equation}

While the boundary condition in Eq.~\eqref{eq: boundary conditions sop gamma bt} is common in streaming problems,  its use for a sharp-edge device is problematic.  Our domain has re-entrant corners at the sharp edges so that the solution of the first-order problem, while bounded, has singular velocity gradients at the sharp edges.  Therefore, Eq.~\eqref{eq: boundary conditions sop gamma bt} implies that, at the sharp edges, not only are the velocity gradients of the second-order solution singular, but the very velocity field is also singular.  This feature of the solution, which is intrinsic to a geometry with re-entrant corners, seems to have been neglected in other studies.  In order to remove the singularity in the second-order velocity, one would have to set the displacement at the sharp edge to zero and, possibly, control its growth away from the edge.  However, this constraint is difficult to justify on practical and physical grounds due to the very design of the microfluidic channel of interest.  Hence, and with the expectation to accurately capture the solution only outside the Stokes boundary layer, we proceeded to determine numerical solutions without taking any special precautions other than the use of a reasonable adaptive mesh refinement scheme as described later.

\subsection{Mean trajectories}
\label{subsec: mean trajectories}
Figure~\ref{fig: ExperimentalTraces} shows the mean trajectories of polystyrene spherical beads placed in the fluid for streaming flow visualization purposes.  As this is the primary piece of experimental evidence at our disposal, we wish to establish whether or not our calculations are able to reproduce a flow pattern similar to that in Fig.~\ref{fig: ExperimentalTraces}.  This is why our code includes an implementation of a tracking strategy proposed by Bruus and co-workers.\cite{Bruus2012Acoustofluidics0,Barnkob2012Physics-of-Micr0,Muller2012A-Numerical-Stu-0,Settnes2012Forces-Acting-o-0,Barnkob2012Acoustic-Radiat-0}  This strategy is predicated on the determination of the radiation force acting on a bead of radius $a$, mass density $\rho_{p}$, and compressibility $\kappa_{p}$ under the influence of a standing wave in the flow.  The bead is modeled as a wave scatterer, and the radiation force is then found to be 
\begin{equation}
\label{eq: radiation force def}
\bv{F}^{\text{rad}} = -\frac{4 \pi a^{3}}{3} \Bigl[
\tfrac{1}{2} f_{1}  \kappa_{0} \nabla \bigl\langle p_{1}^{2} \bigr\rangle
-
\tfrac{3}{4} \rho_{0} \rp(f_{2}) \nabla \langle \bv{v}_{1} \cdot \bv{v}_{1} \rangle
\Bigr],
\end{equation}
where $\kappa_{0} = 1/(\rho_{0} c_{0}^{2})$ is the compressibility of the fluid, $\rp(f_{2})$ is the real part of $f_{2}$, and where
\begin{equation}
\label{eq: f1 and f2 defs}
f_{1} = 1 - \frac{\kappa_{p}}{\kappa_{0}}
\quad \text{and} \quad
f_{2} = \frac{2 (1 - \gamma) (\rho_{p} - \rho_{0})}{2 \rho_{p} + \rho_{0}(1 - 3 \gamma)},
\end{equation}
with
\begin{equation}
\label{eq: gamma delta def}
\gamma = - \tfrac{3}{2} [1 + \I (1 + \tilde{\delta})] \tilde{\delta},\quad
\tilde{\delta} = \frac{\delta}{a},
\quad
\delta = \sqrt{\frac{\mu}{\pi f \rho_{0}}},
\end{equation}
and the symbol `\I' denotes the imaginary unit.  In addition to the radiation force, a bead is assumed to be subject to a drag force proportional to $\bv{v}^{\text{bead}} - \langle \bv{v}_{2} \rangle$, which is the velocity of the bead relative to the streaming velocity.  When wall effects are negligible, the drag force is estimated via the simple formula $\bv{F}^{\text{drag}} = 6 \pi \mu a \bigl(\langle\bv{v}_{2}\rangle - \bv{v}^{\text{bead}}\bigr)$.  The motion of the bead is then predicted via the application of Newton's second law:
\begin{equation}
\label{eq: acc bead}
m_{p} \bv{a}_{p} = \bv{F}^{\text{rad}} + \bv{F}^{\text{drag}},
\end{equation}
where $m_{p}$ and $\bv{a}_{p}$ are the mass and acceleration of the bead respectively.  In many acoustofluidics problems the inertia of the bead can be neglected.\cite{Barnkob2012Physics-of-Micr0}  Doing so, Eq.~\eqref{eq: acc bead} can be solved for $\bv{v}^{\text{bead}}$:
\begin{equation}
\label{eq: bead velocity field}
\bv{v}^{\text{bead}} = \langle\bv{v}_{2}\rangle + \frac{\bv{F}^{\text{rad}}}{6 \pi \mu a}.
\end{equation}
For steady flows, we can identify the bead trajectories with the streamlines of the velocity field $\bv{v}^{\text{bead}}$ in Eq.~\eqref{eq: bead velocity field}.

For an ``ideal tracer,'' a bead with the same density and compressibility as the surrounding fluid, $\bv{F}^{\text{rad}} = \bv{0}$ and the bead's velocity coincides with the streaming velocity.  However, it is well known that the trajectories of the streaming velocity field (or its streamlines in steady problems) are not fully representative of the mean trajectories of the fluid's particles as the latter are subject to a drift effect known as Stokes drift.\cite{Zarembo2013Acoustic-Stream0}  The theory around the Stokes drift is developed without reference to the motion of a bead in the fluid and therefore it can be viewed as a theory for the identification of mean trajectories of fluid particles.  We adopt the theory of Lagrangian mean flow described by B{\"u}hler\cite{Buhler2009Waves-and-Mean-0}, and employed by Vanneste and B{\"u}hler\cite{Vanneste2011Streaming-by-Le0}, in which mean particle paths are the trajectories of a velocity field referred to as the Lagrangian velocity, denoted by $\bv{v}^{\text{L}}$, and given by
\begin{equation}
\label{eq: Lagrangian velocity def}
\bv{v}^{\text{L}} =  \langle\bv{v}_{2}\rangle + \langle (\boldsymbol{\xi}_{1} \cdot \nabla) \bv{v}_{1} \rangle,
\end{equation}
where the field $\boldsymbol{\xi}_{1}(\bv{x},t)$ is the first-order approximation of the lift field $\boldsymbol{\xi}(\bv{x},t)$.  The latter is defined such that $\bv{x} + \boldsymbol{\xi}$ represents the true position at time $t$ of a particle with mean position at $\bv{x}$ (also at time $t$).  By asymptotic expansion, $\boldsymbol{\xi}_{1}$ is such that
\begin{equation}
\label{eq: xi1 def}
\frac{\partial \boldsymbol{\xi}_{1}}{\partial t} = \bv{v}_{1}.
\end{equation}
Equation~\eqref{eq: xi1 def} implies that, once the first-order problem velocity solution of the form $\bv{v}_{1} = \bv{v}_{1}^{c}(\bv{x}) \cos(2 \pi f t) + \bv{v}_{1}^{s}(\bv{x}) \sin(2 \pi f t)$ is computed, $\boldsymbol{\xi}_{1}$ can be calculated during post-processing via an elementary time integration.  For a steady problem, the trajectories of the fluid particles are then the streamlines of $\bv{v}^{\text{L}}$.

Differently from $\bv{v}^{\text{bead}}$, $\bv{v}^{\text{L}}$ is an intrinsic property of the combination of the first- and second-order solutions of the acoustofluidic problem.  That is, $\bv{v}^{\text{L}}$ arises from kinematic arguments alone without reference to the balance of linear momentum or the balance of mass.  As such, the Lagrangian velocity field does not coincide with either $\bv{v}^{\text{bead}}$ or the mean velocity of the mass flow.\cite{Buhler2009Waves-and-Mean-0,Vanneste2011Streaming-by-Le0} The latter, denoted by $\bv{v}^{\text{M}}$, is defined such that $\rho_{0} \bv{v}^{\text{M}}$ gives the second-order approximation of the linear momentum flow per unit volume:
\begin{equation}
\label{eq: vm def}
\bv{v}^{\text{M}} = \langle\bv{v}_{2}\rangle + \frac{1}{\rho_{0}} \langle \rho_{1} \bv{v}_{1} \rangle.
\end{equation}
As already alluded to, $\bv{v}^{\text{bead}}$, $\bv{v}^{\text{L}}$, and $\bv{v}^{\text{M}}$ introduce corresponding notions of mean flow trajectories which can be quite distinct from one another.  However, they all carry useful information about the solution.

\subsection{Numerical solution approach}
\label{subsec: Numerical solution approach}
As is customary in acoustic streaming problems, we seek time-harmonic solution for $\bv{v}_{1}$ and $p_{1}$ in the first-order problem, while we seek steady solutions for $\bv{v}_{2}$ and $p_{2}$ in the second-order problem.  Combining information from these two solutions, it is then possible to estimate the mean trajectory of material particles in the flow.

All the solutions discussed later are for two-dimensional problems.  The numerical solution was obtained via an in-house finite element code based on the \texttt{deal.II} finite element library.\cite{BangerthHartmannKanschat-2007-a,BangerthHartmann-deal.II-Differential--0} For both the first- and second-order problems we used $Q2$-$Q1$ elements for velocity and pressure, respectively, where $Q1$ and $Q2$ denote quadrilateral elements supporting Lagrange polynomials of order one and two, respectively.  Our code was developed using the mathematical framework discussed by K{\"o}ster \cite{Koster2006Numerical-Simul0} who offered a very careful analysis of the numerical properties of the approach.  The main fundamental difference between our code and that by K{\"o}ster is the use of adaptivity.  Specifically, to mitigate somewhat the effects of the singularities discussed earlier, we adopted a very traditional adaptive mesh refinement strategy with an error estimator based on the solution's gradients \cite{Kelly1983A-Posteriori-Er0}.  Our specific error estimator was based on the gradient of the velocity solution of the first-order problem.  The flow patterns we present are those that did not significantly change upon further refinement of the mesh outside the Stokes boundary layer.  Clearly, we make no claims on the values of the velocity gradients within this layer near the sharp edges.

\section{Results and Discussion}
\label{sec: Results and Discussion}

\subsection{Constitutive Parameters}
All of the results presented were obtained using the values in Table \ref{table:constitutive} for the constitutive and operational parameters in the governing equations. Some of the results  pertain to the motion of a dilute concentration of $1.9\umu$ diameter fluorescent beads.
\begin{table}[ht]
\centering
\caption{\label{table:constitutive}Constitutive and operational parameters.} 
\begin{tabular}{l c} 
\toprule
\multicolumn{2}{c}{\textbf{Water}}\\
\cmidrule{1-2}
density ($\rho_{0}$) & $\np[kg\scdot m^{-3}]{1000}$ \\ 
shear viscosity ($\mu$) & $\np[Pa\scdot s]{0.001}$ \\
bulk viscosity ($\mu_{b}$) & $\np[Pa\scdot s]{0.001}$\\
compressibility ($\kappa_{0}$) & $\np[TPa^{-1}]{448}$ \\
speed of sound ($c_{0}$) & $\np[m\scdot s^{-1}]{1500}$ \\
\cmidrule{1-2}
\multicolumn{2}{c}{\textbf{Polystyrene beads}}\\
\cmidrule{1-2}
density ($\rho_{p}$) & $\np[kg\scdot m^{-3}]{1050}$ \\ 
compressibility ($\kappa_{p}$) & $\np[TPa^{-1}]{249}$ \\
diameter ($2 a$) & $1.9\umu$ \\
\cmidrule{1-2}
\multicolumn{2}{c}{\textbf{Operational parameters}}\\
\cmidrule{1-2}
forcing frequency ($f$) & $\np[Hz]{4750}$ \\ 
displacement amplitude ($\|\bv{w}\|_{\Gamma_{t},\Gamma_{b}}$) & $1\umu$ \\ 
\bottomrule
\end{tabular}
\end{table}

\subsection{In \emph{lieu} of convergence tables}
\begin{figure}[htb]
    \centering
    \includegraphics[width=8.5cm]{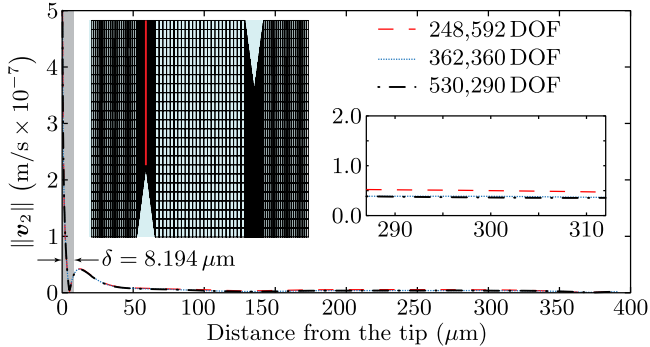}
    \caption{Plot of the magnitude of the second-order velocity vs.\ position along a line parallel to the $y$ axis and emanating from the tip of a sharp edge (red line in left inset). The different curves correspond to different levels of adaptive refinement.  The inset on the left shows the initial mesh used for the calculations.  The right inset shows a zoomed-in portion of the plot outside the Stokes boundary layer of thickness $\delta$.  The units on the vertical axis of the right inset are the same as those on the overall plot.}  
    \label{fig: veconvergence}
\end{figure}

As mentioned earlier, the gradients of the first-order velocity and the second-order velocity field are unbounded at the tips of the sharp edges.  As such, the second order velocity solution does not converge in a strict sense.  Nonetheless, it turns out that the singularity effects are all contained within the Stokes boundary layer and it is therefore still possible to talk about an effective notion of convergence outside this layer.  To illustrate this idea, we present in Fig.~\ref{fig: veconvergence}
plots of the magnitude of the second-order velocity $\bv{v}_{2}$ as a function of position along a line parallel to the $y$ axis (cf.\ Fig.~\ref{fig: periodic cell def}) and emanating from the tip of the lower sharp edge as shown in red in the left inset of Fig.~\ref{fig: veconvergence}.  The latter shows three curves corresponding to increasing adaptive refinement levels identified via the number of \ac{DOF} used in the calculation.  Our calculations have been carried out on increasingly refined meshes using the adaptive refinement strategy described in Sec.~\ref{subsec: Numerical solution approach}.  The left inset depicts the initial (coarsest) mesh, which consists of \np{4032} elements with a total of \np{37251} degrees of freedom.  Our most refined mesh was achieved after nine levels of adaptive refinement and it had \np{1134530} \ac{DOF}.  The curves shown pertain to refinement levels five, six, and seven with \np{248592}, \np{362360}, and \np{530290} degrees of freedom, respectively.  Curves corresponding to further refinement were not plotted because they overlapped the curves shown for refinement levels six and seven.  The right inset, whose vertical axis has units identical to the overall plot, shows a zoomed-in detail of the three curves away from the sharp tip.  To facilitate the discussion, we have used a grey rectangle at the left edge of the plot to distinguish the part of the velocity response contained in the Stokes boundary layer. The thickness of the Stokes layer has been denoted by $\delta$ and computed using the well-known formula $\delta = \sqrt{2\mu/(2 \pi f \rho_{0})}$\cite{Lei2013Acoustic-Stream-0,Lei2014Numerical-Simul-0}.  As expected, the magnitude of $\bv{v}_{2}$ displays an asymptote at the sharp edge.  At the same time, the magnitude of $\bv{v}_{2}$ appears to be well-behaved outside the Stokes layer.

\begin{figure}[htb]
    \centering
    \includegraphics[width=8.5cm]{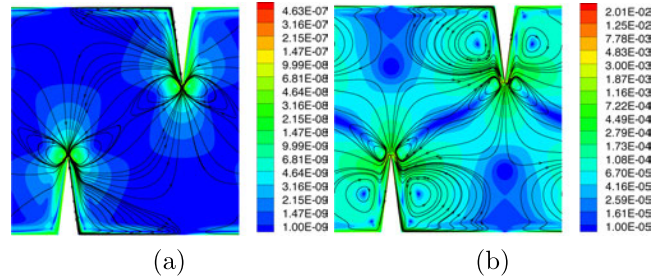}
    \caption{Plots of the velocity $\bv{v}^{\text{bead}}$ in Eq.~\eqref{eq: bead velocity field} corresponding to the boundary conditions in (a) Eq.~\eqref{eq: periodic boundary conditions bt} and (b) Eq.~\eqref{eq: anti periodic boundary conditions tb}.  The color map represents the magnitude of $\bv{v}^{\text{bead}}$ whereas the lines are some of the streamlines of the $\bv{v}^{\text{bead}}$ field. The channel dimensions (cf.\ Fig.~\ref{fig: periodic cell def}) are $L = 300\umu$, $H = 600\umu$, $\alpha = 15^{\circ}$, and $h = 200\umu$. The wall displacement was only in the $y$ direction with magnitude $1\umu$.}
    \label{fig: vbead}
\end{figure}
\begin{figure*}[htb]
    \centering
    \includegraphics{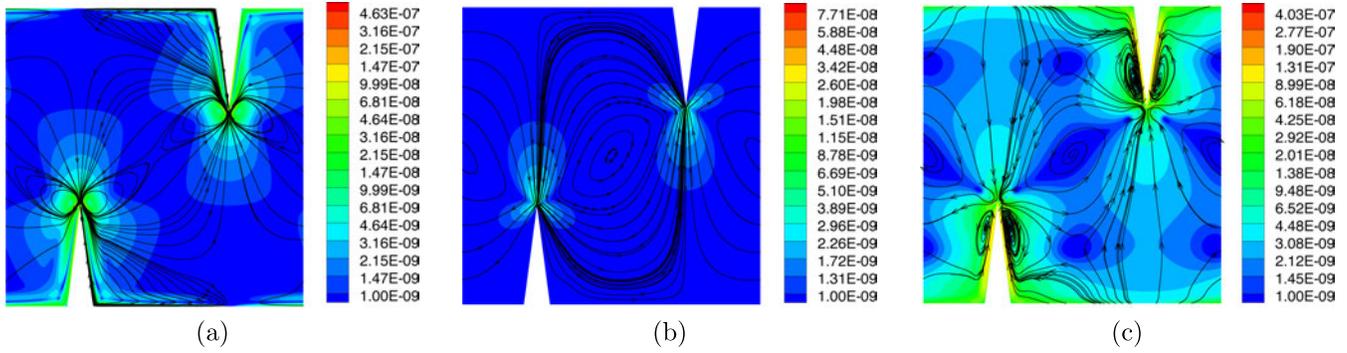}
    \caption{Plots of the mean velocity fields and corresponding streamlines discussed in Section~\ref{subsec: mean trajectories}: (a) $\bv{v}^{\text{bead}}$; (b) $\bv{v}^{\text{L}}$; (c) $\bv{v}^{\text{M}}$.  The color map represents the velocity magnitudes, whereas the lines depict streamlines. In all cases, the channel dimensions (cf.\ Fig.~\ref{fig: periodic cell def}) are $L = 300\umu$, $H = 600\umu$, $\alpha = 15^{\circ}$, and $h = 200\umu$. The wall displacement was only in the $y$ direction with magnitude $1\umu$.}
    \label{fig: MeanTrajectories}
\end{figure*}
Most importantly, the right inset shows that, outside the Stokes layer, $\bv{v}_{2}$ can be considered sufficiently converged after the sixth level of adaptive refinement.  To obtain accurate particle trajectories, all of the results presented in this paper were obtained using refinement levels eight or nine.

\subsection{Effects of boundary conditions on $\Gamma_{t}$ and $\Gamma_{b}$}
In Section~\ref{subsec: loading conditions}, we discussed two possible expressions for the boundary conditions on the $\Gamma_{t}$ and $\Gamma_{b}$ (cf.\ Fig.~\ref{fig: periodic cell def}) portions of the solution's domain.  Figure~\ref{fig: vbead} shows our numerical solution for the velocity $\bv{v}^{\text{bead}}$ in Eq.~\eqref{eq: bead velocity field} of beads in the flow. Referring to Fig.~\ref{fig: periodic cell def}, the results in the above figure were obtained for a channel with $L = 300\umu$, $H = 600\umu$, $\alpha = 15^{\circ}$, and $h = 200\umu$.  In both cases, the wall displacement was completely in the $y$ direction with an amplitude of $1\umu$.  In both cases, we observe that eddies are predicted with streamlines symmetric relative to the (geometric) center of the simulation domain.  This features matches the experimentally obtained bead traces in Fig.~\ref{fig: ExperimentalTraces}.  However, we notice that in the experiments there is no evidence of large eddies at the foot of the sharp edges, as predicted using the anti-periodic boundary conditions in Eq.~\eqref{eq: anti periodic boundary conditions tb}.  Furthermore, we observe that the magnitude of $\bv{v}^{\text{bead}}$ is vastly different in the two cases.  In fact, even in some regions outside the Stokes layer, the velocity predicted in Fig.~\ref{fig: vbead}(b) can be of the same order of magnitude as the average flow velocity observed in experiments with a net flow through the channel (absent in our simulations).  Instead, the magnitudes in Fig.~\ref{fig: vbead}(a) are more compatible with experimental results, as are the predicted streamlines.  As discussed in Section~\ref{subsec: loading conditions}, the boundary conditions used in this work are based on heuristic arguments and under the assumption that the channel is rigid, as opposed to being produced by a more sophisticated piezo-electro-elastic calculation.  Despite this approximation, the result in Fig.~\ref{fig: vbead}(a) is very much in agreement with experimental observations.  

\begin{figure}[htb]
    \centering
    \includegraphics{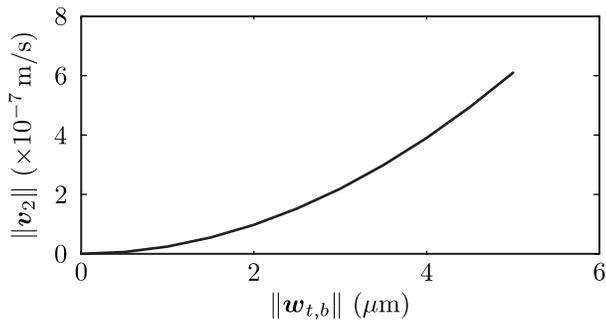}
    \caption{Plot of the magnitude of the second-order velocity as a function of boundary displacement amplitude. $\|\bv{v}_{2}\|$ was measured at a point lying on a line emanating from the tip of a sharp edge parallel to the $y$ direction (cf.\ Fig.~\ref{fig: periodic cell def}) and a distance away from the tip equal to the thickness of the Stokes boundary layer.  The geometry parameters have been kept the same as in preceding calculations.}  
    \label{fig: wamplitude}
\end{figure}
Therefore, for the remainder of the paper we will use only the boundary conditions in Eq.~\eqref{eq: periodic boundary conditions bt}.

\subsection{Mean flows and trajectories}

\begin{figure*}[htb]
    \centering
    \includegraphics{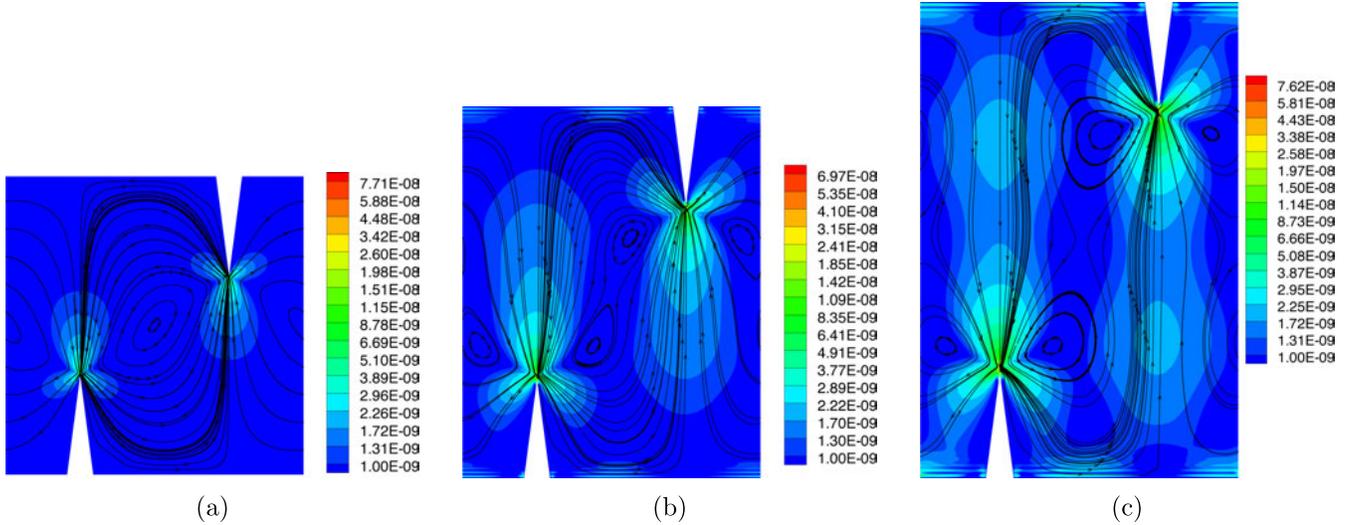}
    \caption{Plots particle trajectories for three different cases with fixed value of $h = 200\umu$ and values of (a) $H = 600\umu$, (b) $750\umu$, and (c) $900\umu$ (cf.\ Fig~\ref{fig: periodic cell def}).  The colormap describes the values of the magnitude of the Lagrangian velocity $\bv{v}^{\text{L}}$, whereas the lines identify the streamlines of $\bv{v}^{\text{L}}$.}
\label{fig: height ratio}
\end{figure*}

Next, we compare the three different types of mean velocities introduced in Section~\ref{subsec: mean trajectories}, along with their corresponding streamlines. Figure~\ref{fig: MeanTrajectories}
shows plots of $\bv{v}^{\text{bead}}$, $\bv{v}^{\text{L}}$, and $\bv{v}^{\text{M}}$ for the boundary conditions given by  Eq.~\eqref{eq: periodic boundary conditions bt}.  Again, the channel has the following dimensions: $L = 300\umu$, $H = 600\umu$, $\alpha = 15^{\circ}$, and $h = 200\umu$; also the wall displacement was completely in the $y$ direction with an amplitude of $1\umu$.  It is important to note that the three plots in Fig.~\ref{fig: MeanTrajectories} are the outcome of a single calculation, that is, they are the product of a single set of geometric parameters and boundary conditions.  As discussed earlier in the paper, we view $\bv{v}^{\text{bead}}$ as the velocity of tracing beads in a fluid. The field $\bv{v}^{\text{bead}}$ is an attempt to capture in an approximate sense the interaction between the beads and the fluid in which they are immersed. Instead, $\bv{v}^{\text{L}}$, which is the sum of the second-order velocity and the Stokes drift term, is the so-called Lagrangian velocity and it is the vector field whose trajectories are the \emph{mean trajectories} of the fluid particles.\cite{Buhler2009Waves-and-Mean-0,Vanneste2011Streaming-by-Le0}. Finally, $\bv{v}^{\text{M}}$ is the average mass flow.  What Fig.~\ref{fig: MeanTrajectories} is meant to show is that the above notions of mean flow can indeed be quite different and therefore represent very distinct properties of the same underlying solution. The difference between the streamlines of $\bv{v}^{\text{L}}$  and the experimental bead trajectories is to be expected. In fact, $\bv{v}^{\text{L}}$ indicates the trajectory of fluid particles in the absence of the beads in the flow, and thus should not be used for comparison with bead trajectories. On the other hand, $\bv{v}^{\text{bead}}$, resulting from the balance of radiation force and Stokes drag, describes the motion of the beads and is indicative of the bead trajectories observed in the experiments. We have already established that the bead trajectories are in good qualitative agreement with the experiments. Therefore, we can use the other measures of mean flow with the same degree of confidence. It may be noted from  Eq.~\eqref{eq: bead velocity field} that the difference between $\bv{v}^{\text{bead}}$ and the second-order velocity depends on a term that is proportional to the square of the radius of the bead. Thus, as the bead size approaches zero, $\bv{v}^{\text{bead}}$ tends to the second-order velocity and the bead tracking method essentially consists in studying the mean trajectory of the second-order velocity solution. 
As the velocity field $\bv{v}^{\text{L}}$ is an intrinsic property of the flow without any beads in it, we feel that it is a more appropriate descriptor of mean fluid particle trajectories and, as such, a more meaningful descriptor of the mixing properties of the sharp-edge device.  Hence, in the remainder of the paper, we will base most of our discussion on plots of $\bv{v}^{\text{L}}$.  Before proceeding further,  we observe that the streamlines of $\bv{v}^{\text{L}}$ do not show eddies for the geometry and boundary conditions used to generate Fig.~\ref{fig: MeanTrajectories}.  This is because, for the stated simulation conditions, the Stokes drift effectively cancels the streaming velocity $\bv{v}_{2}$.  As will be shown later, for other simulation conditions, the Lagrangian velocity will show the existence of eddies in the mean flow.

\subsection{Effect of displacement amplitude}
Next, we study the effect of displacement amplitude prescribed on $\Gamma_{t}$ and $\Gamma_{b}$ on the magnitude of the resulting streaming velocity ($\bv{v}_{2}$).  We simulated the acoustic streaming for different values of the input displacement amplitude and the same geometric parameters used thus far.  
 With this in mind, Fig.~\ref{fig: wamplitude} shows the plot of the magnitude of the second-order (streaming) velocity measured at a point lying on a line emanating from the tip of a sharp edge parallel to the $y$ direction (cf.\ Fig.~\ref{fig: periodic cell def}) and a distance away from the tip equal to the thickness of the Stokes boundary layer. The second-order streaming velocity was found to increase quadratically with displacement amplitude. This is expected since the first-order pressure and velocity depend linearly on the displacement amplitude; and the second-order streaming velocity, in turn, depends quadratically on the first-order pressure and velocity. In the experiments, the amplitude of the acoustic wave is proportional to the square-root of the signal power, for small values of the signal power. Thus, we expect the streaming velocity to be linearly dependent on the input signal power.

\subsection{Effect of the channel dimensions $h$ and $H$}
\begin{figure*}[htb]
    \centering
    \includegraphics{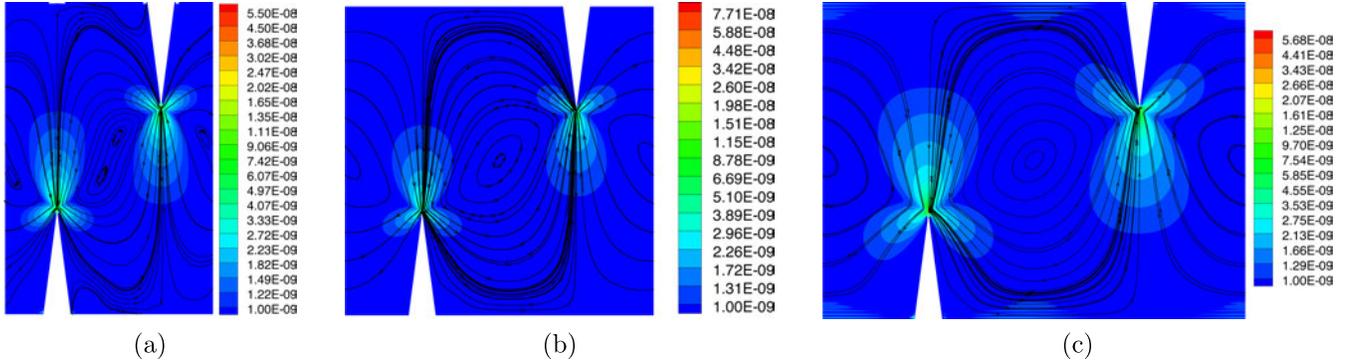}
    \caption{Plots particle trajectories for three different cases with fixed values of $h = 200\umu$ and $H = 600\umu$, and different values of (a) $L = 200\umu$, (b) $300\umu$, and (c) $400\umu$ (cf.\ Fig~\ref{fig: periodic cell def}).  The colormap describes the values of the magnitude of the Lagrangian velocity $\bv{v}^{\text{L}}$, whereas the lines identify the streamlines of $\bv{v}^{\text{L}}$.}
\label{fig: Ldimension}
\end{figure*}

With this section, we turn to an assessment of the effectiveness of sharp edges in acoustofluidic mixing.  Mixing characterization is an inherently complex subject because mixing is a time dependent process.  The assessment offered in this paper is limited to the theoretical modeling employed: the equations presented earlier are for a single flow. Therefore, our conclusions are strictly applicable only to perfectly miscible fluids of same density and constitutive properties.  As important is the fact that we focus on the analysis of steady state acoustic streaming and therefore we do not consider the time evolution of the fluid.  This is not to say that our predictions are inadequate.  In fact, analyzing the structure of steady state mean particle trajectories is analogous to analyzing a dynamical system's response in phase-space to determine the possible evolution of the system as a function of initial conditions, where the latter, in the present context, are the initial positions of fluid particles in the channel.  Therefore, given fluid particles initially distributed as shown in Fig.~\ref{fig: Schematic}(a) near the fluid inlets, we can tell whether or not particles are forced by the device to travel away from their initial position.  While we do not solve a time dependent system, we do solve for the streaming velocity.  This indicator will allow us to assess whether a particular configuration will achieve mixing more rapidly than another.  The mean trajectories we study are the streamlines of the Lagrangian velocity field $\bv{v}^{\text{L}}$.

Referring to Fig.~\ref{fig: periodic cell def}, we begin our analysis by considering the effect of the channel dimensions $h$ and $H$.  We consider three different cases with constant tip size $h = 200\umu$ and the following values for $H$: $600\umu$, $750\umu$, and $900\umu$. In all cases, the separation between sharp edges $L$ was $300\umu$ and the tip angle $\alpha$ was $15^{\circ}$.  It can be seen from Fig.~\ref{fig: height ratio}, no matter the ratio $h/H$, fluid particles near the walls are forced by the device to travel towards the opposite wall.  Also, as the ratio $h/H$ decreases, the flow pattern ``breaks'' into two distinct vortices, thus acquiring the sort of pattern displayed in Fig.~\ref{fig: vbead}.  This result indicates that there is a critical value of $h/H$ for which eddies might not be present.  Unfortunately, we do not have experimental confirmation of this finding at this time.  Nonetheless, this finding points to the conclusion that narrowing channels might not be necessarily desirable.  We view the presence of eddies near the oscillating sharp tips as a mixing enhancer.  This feature indicates that there are trajectories spanning the entire width of the channel in close proximity with trajectories a of very different type, namely with local circulation. While this pattern is certainly not turbulent or chaotic, it does indicate the possibility of good mixing conditions in that a small local random perturbation
\begin{figure}[htb]
    \centering
    \includegraphics[scale=1]{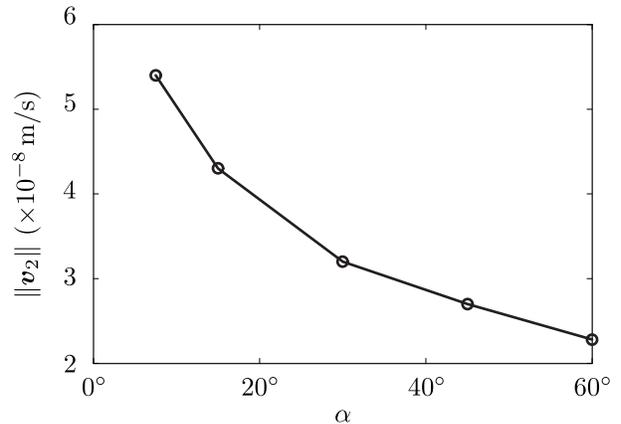}
    \caption{Plot of $\|\bv{v}_{2}\|$ values at a fixed point ahead of a sharp edge vs.\ tip angle $\alpha$ (cf.\ Fig.~\ref{fig: periodic cell def}).  The dimensions of the channel are $h = 200\umu$, $H = 600\umu$ and $L = 300\umu$.  Open circles denote computed data points corresponding to $\alpha = 7.5^{\circ}$, $15^{\circ}$, $30^{\circ}$, $45^{\circ}$, and $60^{\circ}$.  In all cases, $\|\bv{v}_{2}\|$ was calculated at the point lying on a line emanating from the tip of a sharp edge parallel to the $y$ direction and a distance away from the tip equal to twice thickness of the Stokes boundary layer.}
    \label{fig: v2vsalpha}
\end{figure}
 around the oscillating tips can cause particles coming from, say, the top region of the domain to be ``trapped'' (at least temporarily) in a completely different region of the channel. From the viewpoint of mixing with an underlying input flow, the fact that the streaming flow solution displays a very distinct central symmetry indicates that better mixing conditions are achieved by ensuring that the inlet flow be not centered within the channel.  Finally, we observe that, as $H$ increases, the fraction of the solution domain experiencing mid-range velocity magnitudes appears to increase.
\begin{figure*}[htb]
    \centering
    \includegraphics{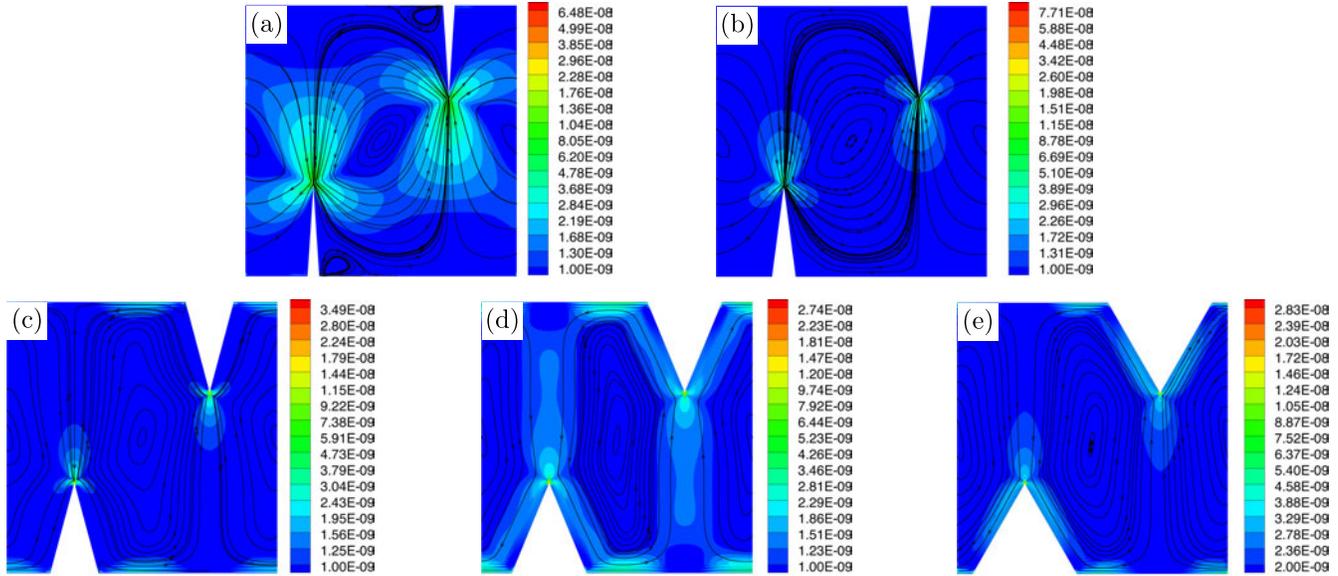}
    \caption{Plot of $\|\bv{v}^{\text{L}}\|$ (colormap) and it streamlines for various values of tip angle $\alpha$ (cf.\ Fig.~\ref{fig: periodic cell def}).  The dimensions of the channel are $h = 200\umu$, $H = 600\umu$ and $L = 300\umu$.  The value of $\alpha$ are (a) $7.5^{\circ}$, (b) $15^{\circ}$, (c) $30^{\circ}$, (d) $45^{\circ}$, and (e) $60^{\circ}$.}
\label{fig: alphatrajectories}
\end{figure*}

\subsection{Effect of the channel dimension $L$}

Referring to Fig.~\ref{fig: periodic cell def}, $L$ is the distance separating sharp edges on opposite sides of the channel walls.  Fig.~\ref{fig: Ldimension} shows the Lagrange velocity streamlines and magnitudes for three different values of the parameter $L$ equal to $200\umu$, $300\umu$, and $400\umu$.  The values of $h$ and $H$ were $200\umu$ and $600\umu$, respectively.  The tip angle was set to $15^{\circ}$.  These results indicate that increasing the distance between opposing sharp edges does not favor the presence of eddies near the tips.  However, the streamlines in Fig.~\ref{fig: Ldimension}(a) seem to indicate the presence of possible stagnation points at the feet of the sharp edges, clearly a feature that would be undesirable.  As far as the magnitude of the velocity is concerned, the figures do not indicate sufficiently strong trends in this regard.

\subsection{Effect of tip angle}

Normally, the mixing of fluids occurs as these fluids move along the channel.  That is, when there is a net flow, governed by the conditions at the inlets, it interacts with the streaming flow.  Provided that we did not consider this interaction, the ability of the device to force particles along the trajectories illustrated thus far depends on the strength of the streaming flow in relation to the background net motion along the channel.  For this reason, it is important to determine which, if any, is the design feature that most decisively contributes to the magnitude of the streaming flow.  We believe this feature to be the angle $\alpha$ (cf.\ Fig.~\ref{fig: periodic cell def}) at the tip of the sharp edge.  It is because $\alpha$ is acute that the second-order solution is singular.  It is therefore important to understand how $\alpha$ affects the strength of the streaming flow.  Normally, one would have to characterize the type of singularity induced by the tip angle and make inferences on the overall strength of the streaming flow. Unfortunately, no analytical results are available on the analysis of the strength of the singularity at re-entrant corners for asymptotic expansions of the compressible Navier-Stokes equations as employed here.  However, in the equations for both the first- and second-order problems, we see the presence of differential operators that are strongly reminiscent of the Navier equations for the linear elastic boundary value problem.  

When a re-entrant corner is present in elastic problems, the displacement gradient experiences singularities of the type $1/r^{\lambda}$, where $r$ is the distance from the tip of the corner and $0< \lambda < \tfrac{1}{2}$.\cite{Sadd2009Elasticity:-The0}  For $\alpha \to 0$, $\lambda \to \tfrac{1}{2}$, which represents the stress/strain behavior found at the tip of a sharp crack.  While, the equations used herein are not identical to those of linear elasto-statics, we speculate that the analogy with fracture problems in elasticity might be an appropriate tool to guide us in the interpretation of the results. In order to quantify how the velocity depends on the tip angle, we have measured the magnitude of $\bv{v}_{2}$ at a distance equal to $2 \delta$ ahead of the tip along a line emanating from the tip and parallel to the $y$, where $\delta$ is the thickness of the Stokes layer.  The calculations are reported in Fig.~\ref{fig: v2vsalpha}.
Clearly, the choice of location at which $\|\bv{v}_{2}\|$ is measured is arbitrary but it is motivated by the fact that $\|\bv{v}_{2}\|$ becomes unbounded as the tip is approached and therefore its measure becomes meaningless.  The plot shows that $\|\bv{v}_{2}\|$ increases with a decrease of tip angle and that the rate of increase also increases as $\alpha$ becomes smaller.  Hence, one immediate conclusion is that the smaller the value of $\alpha$, the stronger the effect of the streaming flow on the overall flow in the device and the better its mixing properties.  However, this conclusion needs to be tested by considering the effect of tip angle on particle trajectories.  This effect was captured in Fig.~\ref{fig: alphatrajectories} for the values of tip angles already mentioned.  What is important to notice is that, for $\alpha = 7.5^{\circ}$ the simulation predicts the appearance of recirculation areas at the feet of the sharp edges that may trap fluid particles permanently.  This effect is not entirely surprising since the angle at the sharp edge feet goes to $90^{\circ}$ as $\alpha$ goes to zero, thus producing stagnation zones in the channel with adverse effect on the device mixing properties.  At the same time, we notice that, as the angle becomes smaller, higher values of particle velocity are present over a larger portion of the solution domain.  The importance of this observation lies in the fact that higher particle velocities can strongly reduce mixing times and therefore have a very enhancing effect on the mixing properties of the device as a whole.

\section{Conclusion}
\label{sec: Conclusion}
We studied the flow around acoustically actuated oscillating sharp edges inside a microchannel using a perturbation approach. The numerical results were compared with experimental results and a very good agreement was observed between them, especially in view of the strong simplifying assumptions adopted in choosing boundary conditions. We demonstrated that a computational domain with periodic boundary conditions can be used to model the full device, resulting in significant savings in computational costs and time. The predicted flow profiles were found to reflect the inherent nonlinearity of the acoustic streaming phenomena as the various patterns identified are not linear scalings of one another. The flow field was found to be heavily dependent on the geometrical parameters of the device like the sharp edge tip angle and the ratio $h/H$ between the distance of the sharp edges from the wall and the overall channel's width. The streaming velocity was also observed to show a quadratic dependence on the applied input displacement and a nonlinear increase with the decrease in tip angle.  At the same time, we showed that properties contributing to the overall mixing effectiveness of the device can be in ``competition'' with each other, making the identification of optimal geometric and working configurations nontrivial.  For this reason, we believe that our computational effort, in addition to providing better understanding of flow around sharp edges in confined microchannels, is also very useful in design optimization of sharp-edge micro-mixers. The latter have numerous applications in many lab-on-a chip processes like biomedical diagnostics, drug delivery, chemical synthesis, enzyme reactions.

\section{Acknowledgments}
\label{sec: Acknowledgments}
We gratefully acknowledge financial support from the National Institutes of Health (Director's New Innovator Award, 1DP2OD007209-01), American Asthma Foundation (AAF) Scholar Award, and the Penn State Center for Nanoscale Science (MRSEC) under grant DMR-0820404.

\bibliographystyle{plain}
\bibliography{NamaEtAl-SharpEdge-12Sept2014}
\end{document}